\documentclass[final,3p,times,twocolumn]{elsarticle}
\usepackage{graphicx}
\usepackage{amssymb}
\usepackage{amsmath}
\usepackage{mathtools}
\usepackage[squaren,Gray]{SIunits}
\usepackage{subfigure}
\usepackage{multirow}
\biboptions{comma,square,sort&compress}
\journal{Computational Materials Science}

\begin{document}

\begin{frontmatter}

\title{Verification of first-principles codes: comparison of total energies, phonon frequencies,  electron-phonon coupling and zero-point motion correction to the gap between ABINIT and QE/Yambo}

\author[ucl]{S. Ponc\'e}
\ead{samuel.ponce@uclouvain.be}
\author[montreal]{G. Antonius}
\author[grenoble]{P. Boulanger}
\author[ILL]{E. Cannuccia}
\author[urome]{A. Marini}
\author[montreal]{M. C\^ot\'e}
\author[ucl]{X. Gonze}

\address[ucl]{Universit\'e catholique de Louvain, Institute of Condensed Matter and Nanosciences,
NAPS Chemin des \'etoiles 8, bte L07.03.01, B-1348 Louvain-la-neuve, Belgium}
\address[montreal]{D\'epartement de Physique, Universit\'e de Montreal, C.P. 6128, Succursale Centre-Ville, Montreal, Canada H3C 3J7}
\address[grenoble]{Institut N\'eel, 25 avenue des Martyrs, BP 166, 38042 Grenoble cedex 9, France}
\address[ILL]{Institut Laue Langevin BP 156 38042 Grenoble, France}
\address[urome]{Consiglio Nazionale delle Ricerche (CNR),Via Salaria Km 29.3, CP 10, 00016, Monterotondo Stazione, Italy}

\begin{abstract}
  With the ever-increasing sophistication of codes, the verification of the implementation of advanced theoretical formalisms becomes critical. In particular, cross comparison between different codes provides 
a strong hint in favor of the correctness of the implementations, and a measure of the (hopefully small) possible numerical differences. We lead a rigorous and careful study of the quantities that enter in the calculation of the zero-point motion renormalization of the direct band gap of diamond due to electron-phonon coupling, starting from the total energy, and going through the computation of phonon frequencies and electron-phonon matrix elements. We rely on two independent implementations : Quantum Espresso + Yambo and ABINIT. We provide the order of magnitude of the numerical discrepancies between the codes, that are present for the different quantities: less than $10^{-5}$~Hartree per atom on the total energy (-5.722~Ha/at), less than 0.07~cm$^{-1}$ on the $\Gamma,L,X$ phonon frequencies (555 to 1330~cm$^{-1}$), less than 0.5\% on the square of the electron-phonon matrix elements and less than 4~meV on the zero-point motion renormalization of each eigenenergies (44 to 264~meV). Within our 
approximations, the DFT converged direct band gap renormalization in diamond due to the electron-phonon coupling is -0.409~eV (reduction of the band gap).
\end{abstract}


\end{frontmatter}

\section{Introduction}
\label{Introduction}

First-principles electronic-structure codes develop and evolve : they adapt to increasing computational capabilities and also include new formalisms, approximations, and numerical methods. 
In addition to the validation of new formalisms and approximations, the verification of implementations is of utmost importance if one wants to deliver reliable new results or compare them to existing ones.
This concern has been the subject of increased attention in the recent years, as witnessed by the set up of ESTEST, a framework for the validation and verification of electronic structure codes \cite{Yuan2010}, and the organization of several related activities under the auspices of the ``Centre Europ\'een de Calcul Atomique et Mol\'eculaire" (CECAM) \cite{CECAM}.

In particular, the first-principle computation of electronic properties, quasiparticles band structures and optical spectra of crystalline solids has reached an unprecedented level of sophistication. Many-body $GW$ calculations \cite{Aulbur1999}, dynamical-mean-field theory  \cite{Georges1996} and Bethe-Salpeter \cite{Onida2002} calculations, that includes excitonic effects, sometimes claim to agree with experimental data at the level of 0.1-0.2~eV. However, the influence of lattice vibrations on electronic properties is usually neglected because it is assumed to lead only to minor corrections, on the order of a few tens of meV. Actually, as reviewed in Ref.~\cite{Cardona2005a}, for materials that contain light atoms like diamond, the inclusion of the influence of lattice vibration is non-negligible, since the renormalization is larger than the claimed accuracy of quasiparticle methods.

For the case of diamond, the closing of the electronic gap has been measured experimentally at different temperatures, and an Einstein oscillator fit has been used to extrapolate the data at zero Kelvin \cite{Cardona2005}, giving a value of 0.37~eV for the renormalization of the indirect band-gap due to the zero-point motion renormalization (ZPR) of atoms. The temperature dependence of the direct band-gap of diamond was also studied experimentally \cite{Logothetidis1992}.

The direct band-gap renormalization has been studied from first-principles approaches. Ram\'irez obtained 0.7 eV using path-integral Monte Carlo simulations \cite{Ramirez2006} and more recently Giustino \emph{et al.}~\cite{Giustino2010} obtained a value of 0.615~eV using the Allen-Heine-Cardona (AHC)~\cite{Allen1976,Allen1981} theory, with the Local Density Approximation (LDA)~\cite{Ceperley1980,Perdew1981} of Density Functional Theory (DFT)~\cite{Martin2004}, a basis of plane waves, and norm-conserving pseudopotentials~\cite{Martin2004}.

However, the first-principle computation of such quantity is particularly delicate, because it is the outcome of several layers of consecutive first-principle calculations : computation of the total energy (and associated relaxation of cell geometry), computation of the phonon frequencies and eigenvectors, computation of the electron-phonon (EP) coupling, and finally, computation of the zero-point motion effect. Not only the choice of a mathematical formalism, with associated approximations (like the above-mentioned Monte Carlo versus DFT possibility), might deliver different values, but the implementation of one well-defined mathematical formalism, with given approximations, needs to be carefully verified. 

At variance with the 0.615~eV result of Giustino {\it et al.}~\cite{Giustino2010}, calculations made by us lead  to a smaller value, on the order of 0.4~eV, on the basis of the implementation partly described in Ref.~\cite{Gonze2011}.
However, the mathematical formalism and numerical approximations were, to our understanding, equivalent to that of Ref.~\cite{Giustino2010}.
This raised the question on whether the accumulation of layers of calculations could yield numerical errors that are as large as 0.2~eV, or whether there might be a problem in the implementations.

In this work, we present a rigorous and careful study of all the quantities that enters into the calculation of the ZPR of the direct band gap of diamond due to EP coupling, on the basis of two different implementations, and provide the values of the numerical discrepancies.  
We work within the AHC formalism with exactly the same numerical approximations, as implemented in ABINIT~\cite{Gonze2009}, on one side, and in Yambo~\cite{Marini2009} on top of Quantum Espresso (QE)~\cite{Giannozzi2009}, on the other side. These implementations have been done completely independently by two different groups.
The ABINIT implementation has been used earlier to study zero-point motion effects on the electronic structure in the above-mentioned Ref.~\cite{Gonze2011}, while the 
YAMBO+QE implementation has been used, independently, in Ref.~\cite{Marini2008,Cannuccia2011,Cannuccia2012,Cannuccia2013}. Unfortunately, we did not
have access to the code used by Giustino {\it et al.}~\cite{Giustino2010}.

We found only small numerical discrepancies between the ABINIT and QE+YAMBO results: 
less than $10^{-5}Ha/at$ on the total energy, 0.07~cm$^{-1}$ on the phonon frequencies, 0.005 on the electron-phonon matrix elements squared (relative difference), and less than 4~meV on the ZPR. 
Given our choice of formalism, and associated approximations, the numerically converged value for the renormalization of the direct band gap in
diamond due to electron-phonon coupling in the AHC formalism is -0.409~eV (reduction of the band gap), from both implementations. Changing the
pseudopotential can lead to larger differences, in any case not larger than 50~meV.

The structure of the article is as follows.
In section 2, we discuss the mathematical theory used in this work. 
In section 3, we give details about the material studied as well as computational parameters and approximations. In section 4, we 
review the results and discuss their impact. We draw the conclusions in section 5.

\section{Theory and methods}
\label{Theory and methods}

\subsection{Ground-state and phonons}

The decomposition of the total energy differs between ABINIT and QE, such that a comparison of energy components needs to be done with care.
The expression for the total ground-state energy per unit cell of a periodic insulator at zero Kelvin, within DFT is \cite{Martin2004,Gonze1997}:

\begin{multline}\label{total1}
  E_{\text{Total}} = \frac{1}{N_\mathbf{k}}\sum_{\mathbf{k}}\sum_{n}^{\text{occ}}\left\langle n\mathbf{k}\left| \hat{T} + \hat{V}_{\text{psp}}\right| n\mathbf{k}\right\rangle \\
  + E_{\text{Hxc}}+E_{\text{Ew}} + E_{\text{psp-core}},
\end{multline}
where $n$ is the band number, $\mathbf{k}$ the wavevector, $\left|n\mathbf{k}\right\rangle$ represents a Kohn-Sham orbital, $\hat{T}$ the kinetic
energy operator, $\hat{V}_{\text{psp}}$ the operator corresponding to the external potential of the electronic system (composed by a local and a
non-local part when the implementation is based on the pseudopotential concept), the $n$-summation is over the occupied bands and the $\mathbf{k}$-summation over a discretization of the Brillouin zone. $E_{\text{Hxc}}$ is the Hartree and exchange-correlation energy functional of the electronic density (expressed per unit cell), $E_{\text{Ew}}$ is the Ewald energy per unit cell (periodic positively charged particles placed in a negatively charged homogeneous background),
and finally $E_{\text{psp-core}}$ is the pseudo-core energy per unit cell. 
It is also possible to define a one-electron contribution per unit cell as:

\begin{equation}\label{total2}
 E_{\text{One-el}} = \frac{1}{N_\mathbf{k}}\sum_{\mathbf{k}}\sum_{n}^{\text{occ}}\left\langle n\mathbf{k}\left| \hat{T} + \hat{V}_{\text{psp}}\right| n\mathbf{k}\right\rangle + E_{\text{psp-core}}. 
\end{equation}

See the appendix for more details concerning the Ewald energy, the pseudocore energy and the one-electron contribution.

The phonon frequencies and eigenvectors can be obtained from Density Functional Perturbation Theory (DFPT) following Refs.~\cite{Baroni1987,Pavone1993,Gonze1997,Baroni2001,Gonze2005}. With $\tilde{C}_{s\alpha,s'\beta}(\mathbf{q})$ being the interatomic force constant matrix in reciprocal space, the phonon frequencies $\omega_{\mathbf{q}\lambda}$ 
and eigendisplacements $\xi_\alpha(\mathbf{q}\lambda|s)$ are linked by the dynamical equation 

\begin{equation}\label{dynamical_equation}
\sum_{s'\beta}\tilde{C}_{s\alpha,s'\beta}(\mathbf{q})\xi_{\beta}(\mathbf{q}\lambda|s')=
M_s \omega_{\mathbf{q}\lambda}^2 \xi_{\alpha}(\mathbf{q}\lambda|s),
\end{equation}
where $s$ labels the atom in the cell (at position $\tau_s$ and with atomic mass $M_s$) and $\alpha$ is a Cartesian coordinate.
Using the orthonormalisation relation
\begin{equation}\label{orthonormalisation}
 \delta_{\lambda' \lambda} = \sum_{s\alpha} M_s \xi_{\alpha}^*(\mathbf{q}\lambda'|s)
\xi_{\alpha}(\mathbf{q}\lambda|s),
\end{equation}
the eigenfrequencies can also be expressed as
\begin{equation}\label{phonon_freq}
 \omega_{\mathbf{q}\lambda}^2 = \sum_{s\alpha}\sum_{s'\beta}\xi_{\alpha}^*(\mathbf{q}\lambda|s) \tilde{C}_{s\alpha,s'\beta}(\mathbf{q})\xi_{\beta}(\mathbf{q}\lambda|s').
\end{equation}

\subsection{Electron-phonon coupling and zero-point motion renormalisation}

The computation of the \textit{ab initio} temperature dependence implies the calculation of the electron-phonon interaction. Following Ref.~\cite{Cannuccia2013} the first-order electron-phonon matrix elements can be computed thanks to DFPT as

\begin{multline}\label{gkk}
g_{nn'\mathbf{k}}^{\mathbf{q}\lambda} = \sum_{s\alpha}(2M_s\omega_{\mathbf{q}\lambda})^{-1/2}e^{i\mathbf{q}\cdot \tau_s}\times \\
 \left\langle n\mathbf{k}\left| \frac{\partial \hat{V}_{\text{scf}}}{\partial R_{s\alpha}}\right| n'\mathbf{k}-\mathbf{q}\right\rangle\xi_\alpha(\mathbf{q}\lambda|s),
\end{multline}
where $\hat{V}_{\text{scf}}$ is the self-consistent mean potential felt by the electrons (which depends on the atomic positions):

\begin{equation}\label{Vscf}
\hat{V}_{\text{scf}}=\hat{V}_{\text{psp}}+\hat{V}_{\text{Hxc}}.
\end{equation}

The first-order electron-phonon matrix element, $g_{nn'\mathbf{k}}^{\mathbf{q}\lambda}$, that will be referred to as the ``GKK'' matrix element, describes the probability amplitude for an electron to be scattered from $\mathbf{k}$ to $\mathbf{k-q}$, with the emission or the absorption of a phonon with crystalline momentum $\mathbf{q}$ belonging to the phonon branch $\lambda$.

The second-order electron-phonon matrix element is:

\begin{multline}
\Lambda_{nn'\mathbf{k}}^{\mathbf{q}\lambda\mathbf{q}'\lambda'} = \frac{1}{2}\sum_{s,\alpha,\beta}\frac{{\xi_{\alpha}^*(\mathbf{q}\lambda|s)\xi_{\beta}(\mathbf{q}'\lambda'|s)}}{2M_s(\omega_{\mathbf{q}\lambda}\omega_{\mathbf{q}'\lambda'})^{1/2}}\times \\
\left\langle n\mathbf{k} \left|  \frac{\partial^2 \hat{V}_{\text{scf}}^{(s)}}{\partial R_{s\alpha}\partial R_{s\beta}}\right| n'\mathbf{k}-\mathbf{q}-\mathbf{q}'\right\rangle.
\end{multline}

The  AHC theory~\cite{Allen1981} allows one to calculate the temperature-dependent change in the electronic eigenenergies, as well as their zero-point renormalization, as the sum of a Fan~\cite{Fan1950,Fan1951} and a Debye-Waller (DW) self-energy term. These two terms can be deduced from the more general many-body formalism~\cite{Cannuccia2013} as: 

\begin{multline}\label{fan}
  \Sigma_{n\mathbf{k}}^{\text{FAN}}(i\omega ,T) = \frac{1}{N_\mathbf{q}}\sum_{n'\mathbf{q}\lambda}\left|g_{nn'\mathbf{k}}^{\mathbf{q}\lambda} \right|^2\times \\
  \Bigg[\left(2n_{\mathbf{q}\lambda}(T)+1 \right) \frac{(i\omega -\varepsilon_{n'\mathbf{k}-\mathbf{q}}-i0^+)}{(i\omega -\varepsilon_{n'\mathbf{k}-\mathbf{q}}-i0^+)^2-\omega_{\mathbf{q}\lambda}^2}\\
 +\omega_{\mathbf{q}\lambda}\frac{(1-2f_{n'\mathbf{k}-\mathbf{q}}(T))}{(i\omega -\varepsilon_{n'\mathbf{k}-\mathbf{q}}-i0^+)^2-\omega_{\mathbf{q}\lambda}^2}\Bigg],
\end{multline}
and
\begin{equation}\label{dw}
	\Sigma_{n\mathbf{k}}^{\text{DW}}(T) = \frac{1}{N_\mathbf{q}}\sum_{\mathbf{q}\lambda} \Lambda_{nn\mathbf{k}}^{\mathbf{q}\lambda-\mathbf{q}\lambda}(2n_{\mathbf{q}\lambda}(T)+1),
\end{equation}
where $n_{\mathbf{q}\lambda}(T)$ is the Bose-Einstein distribution function for the phonon mode $(\mathbf{q},\lambda)$ at temperature $T$, and $f_{n'\mathbf{k}-\mathbf{q}}(T)$ is the electronic occupation. 

The ZPR of the traditional AHC theory~\cite{Allen1981} is recovered by using the following approximations for the Fan term: $\omega\approx \varepsilon_{n\mathbf{k}}$ (the on-the-mass-shell (OMS) limit), $|\varepsilon_{n\mathbf{k}-\varepsilon_{n'\mathbf{k}-\mathbf{q}}}| \gg \omega_{\mathbf{q}\lambda}$ (the adiabatic limit) and by considering only the real part of the self energy:

\begin{equation}\label{ahc}
 \Delta \varepsilon_{n\mathbf{k}}^{\text{AHC}}(T) = 
\Sigma_{n\mathbf{k}}^{\text{DW}}(T)+
\frac{1}{N_\mathbf{q}}
\sum_{n'\mathbf{q}\lambda}\frac{\left| g_{nn'\mathbf{k}}^{\mathbf{q}\lambda} \right|^2 (2n_{\mathbf{q}\lambda}(T)+1)}{\varepsilon_{n\mathbf{k}}-\varepsilon_{n'\mathbf{k}-\mathbf{q}}}.
\end{equation}

From a practical point of view, the DW term is very difficult to calculate, as one needs access to the second derivative of the self-consistent potential (that is not provided by a DFPT calculation of phonons). 
Making use of the translational invariance (if all atoms are displaced by the same amount in the same direction, all physical quantities should be conserved) \cite{Allen1981}, one can rewrite the DW term as a sum of a diagonal contribution and a non-diagonal one. 
The diagonal Debye-Waller (DDW) contribution is the product of first-order electron-phonon matrix that is easy to calculate \cite{Gonze2011}: 

\begin{multline}\label{ddw}
\Sigma_{n\mathbf{k}}^{\text{DDW}}(T)=-\frac{1}{N_q}\sum_{\mathbf{q}\lambda}\sum_{s,s',\alpha,\beta}\frac{\xi_{\alpha}^*(\mathbf{q}\lambda|s)\xi_{\beta}(-\mathbf{q}\lambda|s)}{4M_s\omega_{\mathbf{q}s}}\times \\
\sum_{n'\neq n}\frac{1}{\varepsilon_{n\mathbf{k}}-\varepsilon_{n'\mathbf{k}}}\left[ \left\langle n\mathbf{k} \left| \frac{\partial \hat{V}_{\text{scf}}}{\partial R_{s'\alpha}}\right| n'\mathbf{k}\right\rangle\left\langle n'\mathbf{k}\left| \frac{\partial \hat{V}_{\text{scf}}}{\partial R_{s\beta}} \right| n\mathbf{k}\right\rangle\right. \\
+\left.\left\langle n\mathbf{k}\left| \frac{\partial \hat{V}_{\text{scf}}}{\partial R_{s\beta}} \right| n'\mathbf{k}\right\rangle\left\langle n'\mathbf{k} \left| \frac{\partial \hat{V}_{\text{scf}}}{\partial R_{s'\alpha}}\right| n\mathbf{k}\right\rangle \right].
\end{multline}

The non-diagonal contribution comes from the modification of the screening due to atomic motion.
By opposition with the case of small molecules \cite{Gonze2011}, the effect of the non-diagonal Debye-Waller term is expected to be small in extended system, thanks to the screening of the periodic lattice. Neglecting it corresponds to the rigid-ion approximation.

From a numerical point of view, the term with an energy denominator in Eq.~\eqref{ahc} is omitted when the difference of eigenenergies is smaller than $10^{-6}$ or is smoothed by introducing a small imaginary component.  

Finally, following Sternheimer~\cite{Sternheimer1954}, one can largely speed up the calculation of the sum over states appearing in the Fan and DDW terms. In that case, they are rewritten in terms of a sum limited to an active space (spanning the occupied state with a few extra bands over the valence band maximum) :

\begin{multline}\label{stern}
-\sum_{n'\neq n}\frac{ \left|n'\mathbf{k}\right\rangle\left\langle n'\mathbf{k}\right|\frac{\partial \hat{V}_{\text{scf}}^{(s)}(\mathbf{r})}{\partial R_{s\alpha}}\left| n \mathbf{k}\right\rangle }{\varepsilon_{n\mathbf{k}}-\varepsilon_{n'\mathbf{k}}} = P_{a^{\perp}}\left|\frac{\partial n \mathbf{k}}{\partial R_{s\alpha}} \right\rangle \\
- \sum_{\substack{n' \leq a^{\perp} \\ n' \neq n}}  \frac{ \left|n'\mathbf{k}\right\rangle\left\langle n'\mathbf{k}\right|\frac{\partial \hat{V}_{\text{scf}}^{(s)}(\mathbf{r})}{\partial R_{s\alpha}}\left| n \mathbf{k}\right\rangle }{\varepsilon_{n\mathbf{k}}-\varepsilon_{n'\mathbf{k}}},
\end{multline} 
with $P_{a^{\perp}}$ the projector over the states whose eigenenergies is above the active space threshold and therefore orthogonal to the active space. The result of such a projection is an outcome of a phonon DFPT calculation, and, as such it is available at no additional cost. 
More informations about this last derivation can be found in Ref.~\cite{Gonze2011}.

\section{Material and calculation}
\label{Material and calculation}

The ABINIT, QE and Yambo software applications are described in Refs.~\cite{Gonze2009,Giannozzi2009,Marini2009}, respectively.

\subsection{Ground-state and phonons}

The calculation of structural properties in this work is based on DFT~\cite{Hohenberg1964,Kohn1965,Martin2004} using the LDA~\cite{Ceperley1980,Perdew1981}. A norm-conserving pseudopotential \cite{Troullier1991} accounts for the core-valence interaction and a plane-wave basis set is then used to expand the electronic wavefunctions.
The pseudopotential was generated using the \texttt{fhi98PP} code \cite{Fuchs1999} with a 1.5 atomic unit cut-off radius for pseudization. The valence electrons of Carbon, treated explicitly in the \textit{ab initio} calculations, are generated for the 2s$^{2}$2p$^{2}$3d$^{0}$ configuration. Quite importantly for the comparison between codes, the same pseudopotential file was used by ABINIT and QE. Moreover this pseudopotential is the same as the one used in Ref.~\cite{Giustino2010}. We refer to this pseudopotential as our ``reference" pseudopotential.

Careful convergence checks (error below 0.5~mHa per atom on the total energy) leads to the use of a 6x6x6 $\Gamma$ centered Monkhorst-Pack k-point sampling \cite{Monkhorst1976} of the Brillouin zone  and an energy cut-off of 30~Hartree for the truncation of the plane wave basis set. The lattice parameter of 6.652~Bohr was obtained by structural relaxation of the diamond system. 

Additional tests were performed to assess the influence of the pseudopotential choice. 
In addition to our ``reference" pseudopotential, we considered five other ones.
We will refer to the first one as \texttt{06-C.LDA.fhi} also generated using the \texttt{fhi98PP} code. It is a Troullier-Martins pseudopotential with the Perdew/Wang~\cite{Perdew1992} parametrization of LDA, an atomic cut-off radius of 1.0247 atomic unit and a maximum angular channel of $l=3$. The second one is the \texttt{6c.pspnc}  Troullier-Martin~\cite{Troullier1991} pseudopotential with a 1.4851 atomic unit cut-off radius and a maximum angular channel of $l=1$. The third one is the \texttt{06-C.GGA.fhi} Troullier-Martin pseudopotential with the GGA Perdew/Burke/Ernzerhof~\cite{Perdew1996} parametrization and a 1.0247 atomic unit cut-off radius. The maximum angular channel used is $l=3$ for this pseudopotential. The required cut-off energy for the truncation of the basis set for those three pseudopotential was also 30 Hartree. The fourth one is the \texttt{6c.4.hgh} Hartwigsen-Goedecker-Hutter pseudopotential~\cite{Hartwigsen1998} with a 1.2284 atomic unit cut-off radius and a maximum angular channel of 
$l=1$. An energy cut-off of 60 Ha was required for this pseudopotential. 
 The last one is the \texttt{C.pz-vbc.UPF} VonBarth-Car pseudopotential with a maximum angular channel of l=1 and an energy cut-off of 45~Ha and 1.5 atomic unit cut-off radius. The lattice parameter of the five additional pseudopotential after structural relaxation were 6.648, 6.694, 6.729, 6.675 and 6.663~Bohr, respectively. All the calculations with these pseudopotential were also done with the 6x6x6 unshifted Monkhorst-Pack k-point grid.

\subsection{Electron-phonon coupling and zero-point motion renormalisation}

In order to converge the ZPR below 1 meV, in the original AHC formulation, around 300 unoccupied bands needs to be explicitly included in the summation present in the Fan and DDW terms for diamond. In contrast, only 12 bands were needed to describe the active space when the Sternheimer re-writing is used.

To avoid high symmetry points that might slow down the convergence study (some EP matrix elements might be zero by symmetry and are not representative
of the discretization of an integral) we computed the ZPR correction on a random q-wavevector grid, as described in Ref.~\cite{Cannuccia2013}. The rate of convergence of homogeneous wavevector grid will also be discussed. The statistical analysis to converge the results is explained in the next section. 
The Sternheimer implementation, which speeds up significantly the calculation, is only present in the ABINIT software. Therefore we did the statistical analysis only with ABINIT.

\section{Results and Discussion}
\label{Results and Discussion}

\subsection{Ground-state and phonons}

We started by comparing DFT ground-state total energies between ABINIT and QE using the same ``reference" norm-conserving pseudopotential and the same numerical parameters (plane wave kinetic energy cut-off and wavevector sampling). 
The total energy in ABINIT and QE is decomposed in different  terms detailed in Eqs.~\eqref{total1} and~\eqref{total2}. The comparison between the terms and the total energy is given in the upper panel of Table~\ref{table:basic1}. 
The agreement is excellent: one gets a discrepancy on the order of $10^{-5}$ Ha/atom between the total energies 
computed using the two codes. The disagreement is even smaller for selected contributions : on the order of $10^{-7}$ Ha/atom for the exchange-correlation and Hartree contributions, and about $10^{-9}$ Ha/atom for the Ewald energy.
We did not try to track down the origin of the total energy discrepancy, the agreement being beyond practical needs.

Table~\ref{table:basic1} also shows the agreement between the two codes on the phonon frequencies at some high symmetry points obtained from the DFPT Eq.~\eqref{phonon_freq}. The agreement is also rather good, with less than 0.07~cm$^{-1}$ differences after imposition of the acoustic sum rule (ASR) at $\Gamma$. The imposition of the ASR is discussed e.g. in Ref.~\cite{Gonze1997a}.  Without the imposition of the ASR, the frequency of acoustic modes at $\Gamma$ are small, but non-negligible :  3.335~cm$^{-1}$ for ABINIT, and 8.832~cm$^{-1}$ for QE. 
Such a variation between codes is however sufficient to lead to significant differences in the absolute value of the Fan and DDW terms computed separately, as we shall see later. Concerning the electronic properties, the nine lower eigenenergies, relative to the top of the valence band at $\Gamma$ are compared for the two codes in Table~\ref{table:eigenenergies}. One can see that there is less than 0.0003~eV differences between the two codes.

\begin{table}[ht]
\begin{center}
\begin{footnotesize}
\begin{tabular}{r r@{.}l r@{.}l  }
\hline
    & \multicolumn{2}{c}{ABINIT 7.3.2} & \multicolumn{2}{c}{ QE 4.0.5} \\
\hline   
  Kinetic energy             &  8&450310501 &\multicolumn{2}{c}{-}  \\
  One-electron energy        & \multicolumn{2}{c}{-} & 4&135925595 \\
  Hartree energy             &  0&943336981 & 0&943337120 \\ 
  XC energy                  & -3&567609861 & -3&567609935 \\
  Ewald energy               &-12&955782342 & -12&955782345 \\
  Psp-core energy            &  0&581222385 & \multicolumn{2}{c}{-} \\
  Loc. psp. energy           & -5&093200787 & \multicolumn{2}{c}{-} \\
  NL psp. energy             &  0&197606844 & \multicolumn{2}{c}{-} \\ 
 \textbf{Total energy  }     & \textbf{-11}&\textbf{4441}16277 & \textbf{-11}&\textbf{4441}29565 \\   
\hline 
  Phonon freq. at $q=\Gamma_{1}$ &  3&335 (x3)   & 8&832 (x3) \\ 
                $q=\Gamma_{25'}$ & 1330&408 (x3) & 1330&428 (x3) \\ 
  with ASR imposed           &  0&000 (x3)     & 0&000 (x3)               \\ 
                             & 1330&403  (x3)& 1330&400 (x3)\\                       
  $q=L_3$    & 555&305 (x2)  & 555&319 (x2)  \\ 
  $q=L_{2'}$                 & 1076&250  & 1076&268  \\
  $q=L_{1}$                  & 1235&429 (x2) & 1235&440 (x2) \\
  $q=L_{3'}$                 & 1273&840 & 1273&860  \\
  $q=X_3$    & 795&900  (x2) & 795&964 (x2)  \\
  $q=X_{1}$  & 1098&461 (x2) & 1098&489 (x2) \\ 
  $q=X_{4}$  & 1224&570 (x2) & 1224&590 (x2) \\
\hline 
\end{tabular}
\caption{Comparison of selected quantities related to the ground state and to phonon calculations, for diamond, with a 6x6x6 unshifted k-point grid and a kinetic energy cutoff of 30~Hartree for the plane wave basis set. The same norm-conserving LDA pseudopotential is used. The lattice parameter is 6.652~Bohr. All the energies are in Hartree, are expressed per cell (two atoms per cell) and the phonon frequencies are in $cm^{-1}$. }
\label{table:basic1}
\end{footnotesize}
\end{center}
\end{table}

\begin{table}[ht]
\begin{center}
\begin{footnotesize}
\begin{tabular}{r r r@{.}l r@{.}l }
\hline
\hline
Sym. & Band  &  \multicolumn{2}{c}{eigenergies in Abinit} & \multicolumn{2}{c}{eigenergies in QE} \\
\hline
$\Gamma_1$     & 1     & -21&7959  & -21&7957 \\
$\Gamma_{25'}$ & 2-3-4 &  0&0000  &   0&0000 \\
$\Gamma_{15}$  & 5-6-7 &  5&6698  &   5&6699 \\
$\Gamma_{2'}$  & 8     & 14&3020  &  14&3023 \\
$\Gamma_1$     & 9     & 19&4714  &  19&4716 \\
\hline 
\hline
\end{tabular}
\caption{Comparison between Abinit and QE of the nine lower eigenenergies in eV, relative to the top of the valence band at $\Gamma$.}
\label{table:eigenenergies}
\end{footnotesize}
\end{center}
\end{table}

\subsection{Electron-phonon coupling and zero-point motion renormalisation}

We now move forward and compare the GKK electron-phonon matrix elements given in Eq.~\eqref{gkk}. This quantity is actually subject to an arbitrary dependence on the phase factors of the wavefunctions, and cannot be compared directly between codes. We have therefore compared the square norm of the GKK (the GKK times its complex conjugate). Such a quantity, termed ``GKK2" is relevant in the present context, since the square of GKK is used to build the ZPR, see Eq.~\eqref{fan}. When wavefunctions are degenerate, we also sum them inside the degenerate space, to remove any arbitrariness. Moreover, to decrease the number of handled data, we sum the GKK2 over the six phonon modes, giving SGKK2.

\begin{figure}[htp]
  \centering
  \subfigure[$\Delta$ between SGKK2$_{\Gamma,\Gamma}$ for 30x30 bands ]
{\includegraphics[scale=0.45]{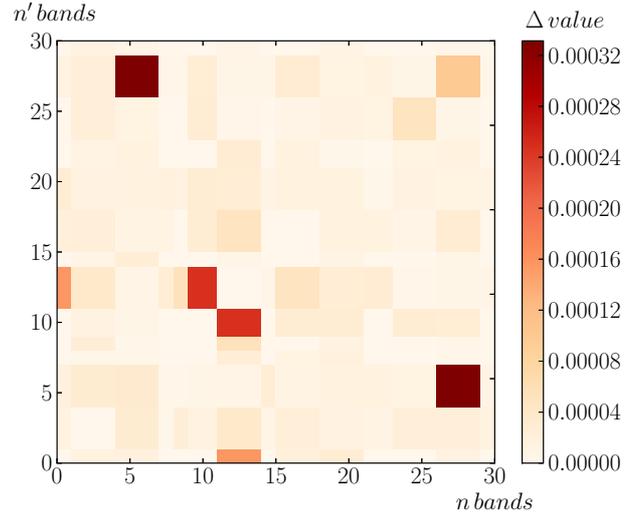}}
   \, 
  \subfigure[$\Delta$ between SGKK2$_{\Gamma,\Gamma}$ for 300x300 bands]
{\includegraphics[scale=0.45]{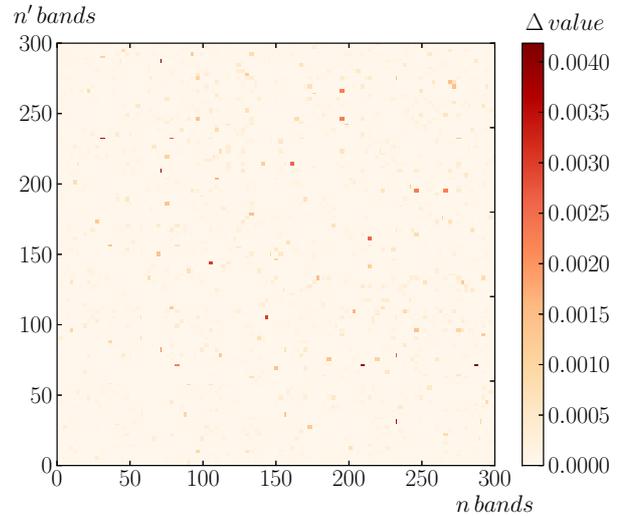}} 
\caption{\label{gkk1} (Color online) Relative differences $\Delta$ of the SGKK2 between ABINIT and QE, at $\mathbf{k}=\Gamma$ and $\mathbf{q}=\Gamma$,  for 20x20 and 300x300 pairs of bands.}  
\end{figure}

\begin{figure}[htp]
  \centering
  \subfigure[$\Delta$ between SGKK2$_{\Gamma,L}$ 300x300 bands]
{\includegraphics[scale=0.45]{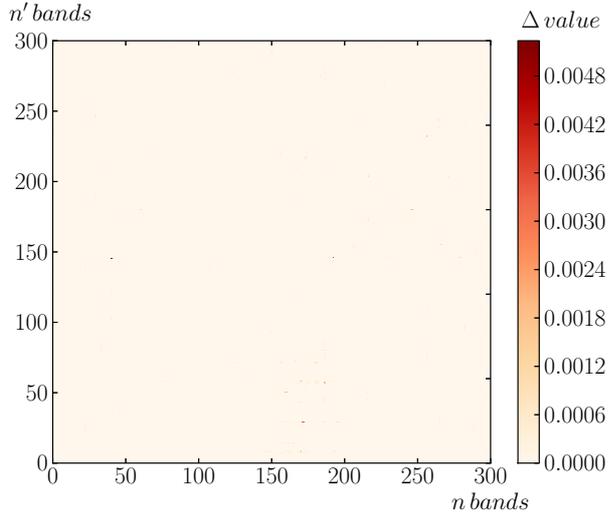}}
  \,
  \subfigure[$\Delta$ between SGKK2$_{\Gamma,X}$ 300x300 bands]
{\includegraphics[scale=0.45]{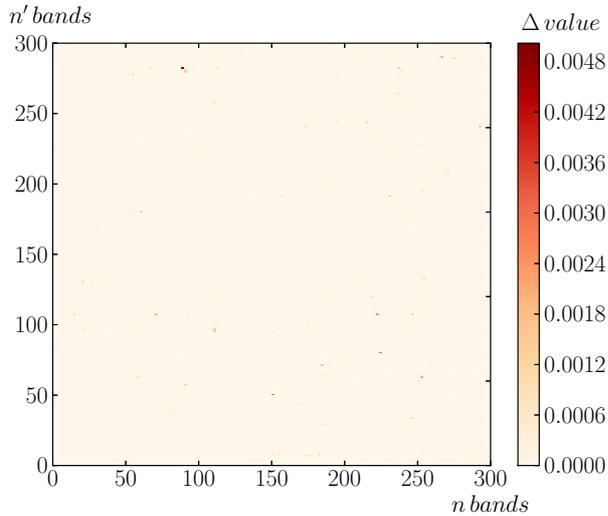}}
  \caption{\label{gkk2} (Color online) Relative differences $\Delta$ of the SGKK2 between ABINIT and QE, at $\mathbf{k}=\Gamma$ and $\mathbf{q}=L$ or $\mathbf{q}=X$ for 300x300 pairs of bands.}  
\end{figure}

A measure of the relative difference between the two codes for SGKK2, for different high symmetry q-wavevectors is displayed in Fig.~\ref{gkk1} and Fig.~\ref{gkk2}. We plot, for each pair of electronic state (or degenerate state) the difference of the SGKK2 divided by their sum :

\begin{equation}\label{Delta}
\Delta= \left| \frac{SGKK2(ABINIT)-SGKK2(QE)}{SGKK2(ABINIT)+SGKK2(QE)} \right|.  
\end{equation}

The absolute values of SGKK2 are reported in Table~\ref{table:gkk_val_abs} for the two codes.

\begin{table}[ht]
\begin{center}
\begin{footnotesize}
\begin{tabular}{r c | r@{.}l | r@{.}l | r@{.}l | r@{.}l | r@{.}l |}
\hline
\hline
 Band & Soft. &  \multicolumn{2}{c}{1} & \multicolumn{2}{c}{2-3-4} & \multicolumn{2}{c}{5-6-7} & \multicolumn{2}{c}{8} & \multicolumn{2}{c}{9} \\
\hline
\multirow{2}{*}{1} & AB      &  \multicolumn{2}{c|}{0}      & \multicolumn{2}{c}{} & \multicolumn{2}{c}{} & \multicolumn{2}{c}{} & \multicolumn{2}{c}{}\\
& QE      &  \multicolumn{2}{c|}{0}      & \multicolumn{2}{c}{} & \multicolumn{2}{c}{} & \multicolumn{2}{c}{} & \multicolumn{2}{c}{}\\
\cline{1-6}
 \multirow{2}{*}{2-3-4} & AB &  1&530449 & 5&803074 & \multicolumn{2}{c}{} & \multicolumn{2}{c}{} & \multicolumn{2}{c}{}\\
& QE &  1&530401 & 5&803088 & \multicolumn{2}{c}{} & \multicolumn{2}{c}{} & \multicolumn{2}{c}{} \\
\cline{1-8}
 \multirow{2}{*}{5-6-7} & AB &  0&493950 & 0&292491 & 4&495984 & \multicolumn{2}{c}{} & \multicolumn{2}{c}{}\\
 & QE &  0&493932 & 0&292472 & 4&496296 & \multicolumn{2}{c}{} & \multicolumn{2}{c}{} \\
\cline{1-10}
\multirow{2}{*}{8} & AB    &  \multicolumn{2}{c|}{0}      & 4&635255 & 2&430641 & \multicolumn{2}{c|}{0} & \multicolumn{2}{c}{}  \\
& QE    &  \multicolumn{2}{c|}{0}      & 4&635284 & 2&430665 & \multicolumn{2}{c|}{0} & \multicolumn{2}{c}{}  \\
\hline
\multirow{2}{*}{9} & AB     &  \multicolumn{2}{c|}{0}      & 1&565383 & 4&002821 & \multicolumn{2}{c|}{0} & \multicolumn{2}{c|}{0} \\
& QE   &  \multicolumn{2}{c|}{0}      & 1&565460 & 4&002843 & \multicolumn{2}{c|}{0} & \multicolumn{2}{c|}{0} \\
\hline 
\hline
\end{tabular}
\caption{Comparison between Abinit (AB) and Quantum Espresso (QE) of absolute value of the SGKK2, at $\mathbf{k}=\Gamma$ and $\mathbf{q}=\Gamma$ in $10^{-6}$~a.u. (1 a.u. = $4.78599\cdot10^{12}J/kg$). Matrix elements with values lower than $10^{-11}$~Ha have been put to 0.}
\label{table:gkk_val_abs}
\end{footnotesize}
\end{center}
\end{table}

One can see that the relative differences are in all three cases lower than 0.005 for all matrix elements on the 300x300 matrix bands.  

Finally, we have compared the ZPR computed with ABINIT and Yambo using the GKKs of QE in the AHC framework of Eq.~\eqref{ahc}.
The energy denominator was smoothened by introducing a small imaginary component of 100~meV, following Ref.~\cite{Zollner1992,Giustino2010}. 

We have first compared the two codes without the Sternheimer rewriting and then, in the case of ABINIT, we have used the Sternheimer rewriting of Eq.~\ref{stern} and we have summed over 300 bands in the case of Yambo. In Table~\ref{table:ahc} we show a comparison between ABINIT and YAMBO for different number of q-wavevectors. 
The 47 q-wavevector case corresponds to a homogeneous, non-shifted 10x10x10 grid, folded in the irreducible part of the Brillouin zone. For the 1000 and 2000 q-wavevector cases, the wavevectors are randomly generated once and then used in both codes. 
In the last two columns of Table~\ref{table:ahc}, we can see that the disparity between the two codes on the ZPR is lower than 4 meV. 

We have imposed in both codes the phonon frequencies to be 0 for the acoustic modes at $\mathbf{q}=\Gamma$ (ASR).
 
One can nevertheless see that the absolute value of the Fan (last term in Eq.~\eqref{ahc}) and DDW (Eq.~\eqref{ddw}) terms display more variation between both codes than the total ZPR (which is the sum of both terms). 
The reason for this is that the acoustic modes tends to have a larger relative difference than the optical ones between the two codes. 
Their separate contributions in the Fan and DDW tends to the same value, with opposite sign, when the limiting behaviour for vanishing wavevector is considered. There is thus a cancellation of error between the Fan and DDW terms, that allows one to obtain a much better accuracy on the sum of these terms. 
Indeed, due to the presence of the phonon frequency in the denominator of Eqs.~\eqref{gkk} and \eqref{ddw}, the acoustic modes will be the one that contributes mostly to the Fan and DDW terms. It can be shown that, due to translational invariance, the eigendisplacement vectors of Fan and DDW will tends to cancel out for acoustic modes (especially those close to $\Gamma$). As a result, mostly the optical modes will contributes to the ZPR. This explains why the discrepancy is larger on the absolute value of Fan and DDW terms separately, than on the total ZPR between the two codes.   
Note that the Fan and DDW terms are not observable quantities separately. They come from a perturbation series, whose sum is an observable.

\begin{table*}[ht]
\begin{center}
\begin{footnotesize}
\begin{tabular}{r r r@{.}l r@{.}l r@{.}l r@{.}l r@{.}l r@{.}l }
\hline
\hline
 &  & \multicolumn{4}{c}{Fan} & \multicolumn{4}{c}{DDW} & \multicolumn{4}{c}{Fan+DDW} \\
Set of q-wavevectors & Band & \multicolumn{2}{c}{ABINIT 7.3.2} &  \multicolumn{2}{c}{Yambo 3.4.0} & \multicolumn{2}{c}{ABINIT 7.3.2} &  \multicolumn{2}{c}{Yambo 3.4.0} & \multicolumn{2}{c}{ABINIT 7.3.2} &  \multicolumn{2}{c}{Yambo 3.4.0} \\ 
& & \multicolumn{2}{c}{SEq / 300 bands} &  \multicolumn{2}{c}{300 bands} & \multicolumn{2}{c}{SEq / 300 bands} &  \multicolumn{2}{c}{300 bands} & \multicolumn{2}{c}{SEq / 300 bands} &  \multicolumn{2}{c}{300 bands} \\ 
\hline
47   &1 & -120&76/-116.54  &  117&30   & 59&23/55.23    &  55&69   &  -61&53/-61.30  & -61&65 \\
& 2-3-4 & -981&61/-969.44  & -978&00   & 1119&92/1107.28  & 1116&53  &  138&30/137.84  &  138&50 \\
& 5-6-7 & -1332&55/-1318.55 & -1329&10  & 1005&15/994.69  & 1002&88  & -327&40/-323.86  & -326&20 \\
& 8     &  -555&40/-541.89 & -543&70   & 60&42/50.32    &  50&76   & -494&98/-491.57  & -492&90 \\
& 9     &  -33&72/-28.50  & -28&49    & -34&89/-39.91   & -40&24   & -68&61/-68.41   &  68&73 \\
\hline  
1000 &1 & -121&13  & -117&70  & 59&46   & 55&90   &  -61&67 & -61&79 \\ 
& 2-3-4 & -983&51  & -979&90  & 1124&21 & 1120&82 &  140&70 &  140&90 \\
& 5-6-7 & -1272&74 & -1269&20 & 1009&01 & 1006&74 & -263&73 & -262&50 \\
& 8     & -284&45  & -272&80  & 60&64   & 50&96   & -223&80 & -221&80 \\
& 9     & -9&83    & -4&55    & -35&03  & -40&39  & -44&85  & -44&95 \\
2000 &1 & -121&20  & -117&80  & 59&45   & 55&90   &  -61&75 & -61&87 \\ 
& 2-3-4 & -983&56  & -980&00  & 1124&11 & 1120&72 & 140&54  &  140&70 \\
& 5-6-7 & -1269&55 & -1266&00 & 1008&92 & 1006&65 & -260&63 & -259&40 \\
& 8     & -293&01  & -281&30  & 60&64   & 50&95   & -232&37 & -230&40 \\
& 9     & -8&83    & -3&56    & -35&02  & -40&39  & -43&86  & -43&95 \\
\hline   
\hline 
\end{tabular}
\caption{Comparison of the ZPR for different electronic states at $\Gamma$, for a 6x6x6 unshifted k-point grid with an energy cutoff of 30~Hartree for the plane wave basis set, using the same norm-conserving LDA pseudopotential. In the case of YAMBO, 300 bands were explicitly included into the calculation. In the case of ABINIT, the Sternheimer equation (SEq) was used to limit the computational effort (12 active bands were needed). Moreover, for the set of 47 q-wavevectors, the value obtained without the Sternheimer equation and with a summation over 300 bands, is also displayed. The energies are in meV.}
\label{table:ahc}
\end{footnotesize}
\end{center}
\end{table*}

\subsection{Analysis of the convergence with respect to the number of q-wavevectors}

We have just provided an analysis of the level of agreement that one can expect from two different codes that implement the same physics. We now turn ourselves to a careful convergence study of the ZPR within the AHC formalism. Since the calculations are heavy in YAMBO due to the band summations we decided to make that convergence study in ABINIT only, with the Sternheimer rewriting.

\begin{figure}[htp]
  \centering
 \includegraphics[scale=0.42]{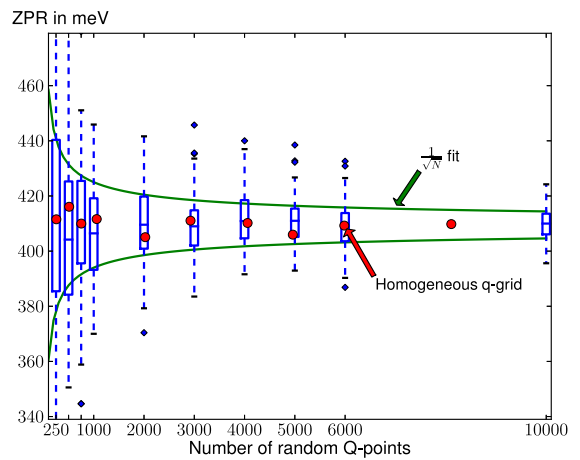}
\caption{\label{Diamond_conv} (Color online) Convergence with respect to the number of random q-wavevetors included in the ZPR calculation using the software ABINIT in the static AHC formulation. 100 ZPR calculations have been performed for each subset of q-wavevectors taking $N_q$ q-wavevectors among 20,000 (the total number of computed q-wavevectors). The upper and lower bars are the maximal and minimal values in each set. The top and bottom of the boxes represent 25\% and 75\% of all the data in the set. The middle line is the median and the blue diamonds are outliers. The red dots comes from non shifted homogeneous Monkhorst-Pack grids, for which the number of q-wavevectors corresponds to those in the irreducible Brillouin zone}  
\end{figure}

We have performed DFPT calculations on 20,000 randomly generated q-wavevectors in the full Brillouin Zone. We have then performed a statistical analysis of these results. We have computed the ZPR over N (N=250, 500, 750, 1000, 2000, 3000, 4000, 5000, 6000, 10000) q-wavevectors taken randomly between the 20,000 set and we have done such calculation 100 times for each N. This gave us, for each N, a set of 100 different ZPR values whose statistical characteristics are given in figure \ref{Diamond_conv}. We can see that the ZPR converges smoothly towards 409~meV, the mean of the ZPR for the 20,000 set.

Since the rate of convergence of the variance of a normal distribution goes as $1/N_q$ with $N_q$ the number of random q-wavevectors, the rate of convergence of the associated standard deviation goes as $1/\sqrt{N_q}$. We can see on Fig.~\ref{Diamond_conv} that the $1/\sqrt{N_q}$ of the continuous line follow neatly the lower 25\% and upper 75\%. 

The drawback of the random q-points methods is that one is forced to test a sufficiently large set of random q-wavevectors. The homogeneous grid approaches might be more appealing. The red dots on Fig.~\ref{Diamond_conv} corresponds to non shifted homogeneous Monkhorst-Pack grids closest to the random points number we have chosen to analyse (e.g. the last grid is a 70x70x70 unshifted q-point grid that lead to 8112 q-wavevectors in the irreducible Brillouin-Zone.). As we can see, the red dots are always well inside the 50\% windows.

One can set an upper limit on the convergence rate if one does not use an imaginary component to smooth the function.

In this case, when the difference of eigenenergies in the denominator of Eq.(11) vanishes, the integrand to be considered over the whole Brillouin zone diverges. 
This happens around $\Gamma$, with a divergence that behaves like $\frac{1}{q^2}$. 
Treating separately a small volume around $\Gamma$, set aside of the regular discretization, one can estimate its contribution by replacing it by the integral over a sphere with a cut-off radius $q_c$ whose length is inversely proportional to the linear density of q-wavevectors.
The contribution of this small sphere is $\propto \int_0^{q_c} \frac{1}{q^2}q^2 dq$
so that the rate of convergence of these integrals goes as $q_c \propto N_q^{-1/3}$, slightly worse than in the case of the random sampling.

Moreover, there are also regions distant from $\Gamma$, but where the eigenenergies of the q-wavevectors are very close to the one at $\Gamma$ (diamond is indeed an indirect gap semiconductor). On the surface $S(\epsilon_{\Gamma})$ where the eigenenergy is exacly equal to the $\Gamma$ eigenenergy, the denominator also vanishes.
In the neighbourhood of the surface, the divergence is inversely proportional to the linear difference between the energy at $\Gamma$ and the actual eigenenergy in its neighbourhood.
To estimate the rate of convergence with respect to the number of q-wavevectors of the discretized integral, we have  to consider the disctretization of an integral in the Brillouin zone, in a zone of width $q_c$ around $S(\epsilon_{\Gamma})$, in which the distance with respect to the surface is denoted as $q_\perp$ giving a behaviour $\propto  S(\epsilon_{\Gamma}) \times \int_{-q_c}^{+q_c} \frac{1}{q_\perp}dq_\perp$.
Although the principal value of this integral vanishes identically, fluctuations due to the discretization will not be small, and hence the convergence is non-monotonic.

Nonetheless, in practice, the small imaginary component at the denominator is present.
One can observe, in Fig. \ref{Diamond_conv}, that the fluctuations, in the case of the homogeneous grid, are quite acceptable.
The error with respect to the q-wavevector sampling might be estimated at 5 meV, for the set of 20,000 q-wavevectors. 

After this careful comparison between codes, and this convergence analysis, we obtain that the 
ZPR converges smoothly towards 409 meV. 

This value disagrees with the one (0.615 eV) provided by Ref.~\cite{Giustino2010}.
The latter was actually first confirmed using QE+YAMBO, see e.g. Ref.~\cite{Cannuccia2012}.  However, while performing the cross verification between ABINIT and QE+YAMBO  for the present study, we found a misuse of the symmetries at $\Gamma$ in the interfacing between QE and YAMBO, affecting only the DDW term. 
After correction, we obtain the results provided in this work, with the numerical uncertainty being much smaller than 0.2 eV. 
Documentation describing how to generate data at $\Gamma$ with the same standard meaning as data at other k-points appeared in QE version later than 4.0.5 (input variable \texttt{nogg}). Work relying on such data might have been affected by this ambiguity.


\subsection{Pseudopotential choices}

We will now assess the influence of the pseudopotential choices. Such a study would not be mandatory 
in the present context of comparison between codes for the same pseudopotential (the reference pseudopotential has indeed been used with ABINIT, YAMBO, and also in the study of Ref.~\cite{Giustino2010}).
This comparison will be performed only using the Abinit software. We have tested all the norm-conserving pseudopotentials available on the Abinit website as well as two UPF pseudopotentials, one of which is the reference pseudopotential.
In the Table \ref{table:psp}, we give a comparison of the ZPR using different pseudopotentials for Carbon. The calculations are made on an homogeneous 10x10x10 $\mathbf{q}$-point grid (47 $\mathbf{q}$-points in the IBZ), for a 6x6x6 unshifted k-point grid with the energy cut-off reported in Section 3.1 for the plane wave basis set, and 12 bands were used (with the Sternheimer equation). The low density q-wavevector grid used in this study allows for a fair comparison between pseudopotentials, but does not yield converged final results.

\begin{table}[ht]
\begin{center}
\begin{footnotesize}
\begin{tabular}{r c r@{.}l r@{.}l r@{.}l c}
\hline
\hline
Pseudo & Band &\multicolumn{2}{c}{Fan} & \multicolumn{2}{c}{DDW} & \multicolumn{2}{c}{Fan+DDW} & \multicolumn{1}{c}{ZPR} \\
\hline
\texttt{\tiny{reference}}         & 2-3-4 & -981&61  &  1119&92 &  138&30 & \multirow{2}{*}{465.70} \\
                      & 5-6-7 & -1332&56 &  1005&16 & -327&40 &   \\
\texttt{\tiny{06-C.LDA.fhi}} & 2-3-4 & -980&90  &  1119&42 &  138&52 & \multirow{2}{*}{467.73} \\
                      & 5-6-7 & -1333&64 &  1004&44 & -329&20 &  \\
\texttt{\tiny{6c.pspnc}}     & 2-3-4 & -938&85  &  1074&14 &  135&28 & \multirow{2}{*}{468.32} \\
                      & 5-6-7 & -1286&04 &  953&00  & -333&03 & \\
\texttt{\tiny{06-C.GGA.fhi}} & 2-3-4 & -952&12  &  1090&20 &  138&09 & \multirow{2}{*}{477.12} \\
                      & 5-6-7 & -1324&85 &  985&82  & -339&03 & \\
\texttt{\tiny{6c.4.hgh}}     & 2-3-4 & -1512&58 &  1649&55 &  136&97 & \multirow{2}{*}{450.35} \\
                      & 5-6-7 & -1791&63 &  1478&25 & -313&38 &  \\
\texttt{\tiny{C.pz-vbc.UPF}} & 2-3-4 & -1027&72 &  1167&13 &  139&41 & \multirow{2}{*}{419.22}\\
                      & 5-6-7 & -1303&42 &  1023&61 & -279&81 & \\
\hline   
\hline 
\end{tabular}
\caption{Comparison inside Abinit of the ZPR (and its Fan + DDW decomposition) for different pseudopotentials at $\Gamma$.
Homogeneous 10x10x10 $\mathbf{q}$-point grid (47 $\mathbf{q}$-points in the IBZ), for a 6x6x6 unshifted k-point grid with the adapted energy cut-off for the plane wave basis set, 12 bands with were used (with the Sternheimer equation). The energies are in meV. Due to the low sampling on the q-wavevector, these value are not converged one, although the comparison between different pseudopotentials is meaningful.}
\label{table:psp}
\end{footnotesize}
\end{center}
\end{table}

One can see that although the pseudopotentials are very different (various exchange-correlation functional, different angular momentum channel include and different atomic cut-off radius) the spread on the ZPR is only around 50 meV. Fluctuations for the Fan and DDW terms, treated separately, are much larger. As emphasized earlier, the decomposition is indeed non-physical, and prone to large numerical uncertainties.

\section{Conclusions}
\label{Conclusions}

In this work, we have carefully compared all the quantities entering into the calculation of the ZPR in the AHC formalism in two different softwares: ABINIT and Yambo on top of QE. We show that one can get less than $10^{-5}$ Ha/atom discrepancy on the total energy,  0.07 cm$^{-1}$ on the phonon frequencies, 0.005 on the electron-phonon matrix elements squared (relative discrepancy) and less than 2 meV on the zero-point motion renormalization. We also discuss the absolute value of the Fan and DDW terms taken separately.
We have also presented the converged result of the band-gap reduction due to electron-phonon renormalization, that is 409 meV at 0 Kelvin and discussed its discrepancy with previously published result. 
We have also performed an analysis of the convergence rate of $\mathbf{q}$-wavevector samplings.

Finally we have discussed the impact of the pseudopotential choices and shown that it was relatively small (around 10\% of the total ZPR) thus increasing our confidence in the results and methodology.

\section{Acknowledgements}
\label{Acknowledgements}
The authors acknowledge many interesting discussions with Giustino and for sharing with us the reference pseudopotential. The authors are also thankful for the scientific and technical help from J.-M. Beuken, A. Jacques, Y. Pouillon, and G.-M. Rignanese.
This work was supported by the FRS-FNRS through a FRIA grant (S.P.). A. M. acknowledges funding by MIUR FIRB Grant No. RBFR12SW0J.
Computational ressources have been provided by the supercomputing facilities of the Universit\'e catholique de Louvain (CISM/UCL) as well as from the Consortium des equipements de Calcul Intensif en F\'ed\'eration Wallonie Bruxelles (CECI) that is funded by the Fonds de la Recherche Scientifique de Belgique (FRS-FNRS).

\section{Appendix}
\label{Appendix}

In this appendix, we detail several terms of the decomposition of the total energy, as provided by ABINIT and QE.
As concern ABINIT, a decomposition of the total energy can be inferred from Refs. \cite{Gonze1997, Gonze1997a}, but the Ewald and psp-core terms are actually mixed in these references, which is misleading.

We define first the psp-core energy.

The external potential originates from sum of atomic pseudopotentials :
\begin{multline}\label{A2}
  v_{\text{psp}}(\mathbf{r},\mathbf{r'}) = \sum_{ls}v_{s}(\mathbf{r}-\mathbf{\tau}_s-\mathbf{R}_l,\mathbf{r}'-\mathbf{\tau}_{s}-\mathbf{R}_l)
\end{multline}
Each atom contribution to this external potential is made of a local and a nonlocal part:

\begin{equation}
  v_{s}(\mathbf{r},\mathbf{r'}) =  v_{s}^{\text{loc}}(\mathbf{r})\delta(\mathbf{r}-\mathbf{r'})+v_{s}^{\text{non-loc}}(\mathbf{r},\mathbf{r}')
\end{equation}

For each atom the local part is long ranged, with an asymptotic behaviour $-Z_s/r$. Such behaviour implies a divergence at $\mathbf{G}=\mathbf{0}$ in reciprocal space. Divergencies at $\mathbf{G}=\mathbf{0}$ also happen in the Hartree energy and the Ewald energy.


A careful treatment of the divergencies lead to their mutual cancellation, albeit with some finite residual.
The residual specifically linked to the long-range behaviour of the local pseudopotential is denoted as the psp-core energy:

\begin{equation}\label{A3}
E_{\text{psp-core}} = \frac{1}{2\Omega_0} \left(\sum_{s}Z_s\right) \sum_{s'} \int \left( v_{s'}^{\text{loc}}(\mathbf{r})+\frac{Z_{s'}}{r}\right)d\mathbf{r}.
\end{equation}

The Ewald energy is the energy of an infinite number of periodic positively charged particle placed in a negative homogeneous background:

\begin{multline}\label{A1}
E_{\text{Ew}} = \frac{1}{2}\sum_{s,s'} Z_{s}Z_{s'}\Bigg[ \sum_{\mathbf{G\ne 0}}\frac{4\pi}{\Omega_0 G^2}e^{i\mathbf{G}\cdot(\tau_{s}-\tau_{s'})}e^{\frac{-G^2}{4\Lambda^2}} \\
- \sum_l \frac{ e^{i\mathbf{q}\cdot \mathbf{R}_l} \text{erfc}(\Lambda \left|\mathbf{R}_l+\tau_{s'}-\tau_s\right|)}{\left|\mathbf{R}_l+\tau_{s'}-\tau_s\right|}-\frac{2}{\sqrt{\pi}}\Lambda \delta_{ss'}-\frac{\pi}{\Omega_0\Lambda^2} \Bigg] 
\end{multline}

with $Z_{s}$ the charge of ion $s$, $\Omega_0$ the unit cell volume, $\Lambda$ a parameter that can assume any value and is adjusted to obtain the fastest convergence.

%

\bibliographystyle{model1a-num-names}
\bibliography{article_v12}

\begin{thebibliography}{41}
\expandafter\ifx\csname natexlab\endcsname\relax\def\natexlab#1{#1}\fi
\providecommand{\url}[1]{\texttt{#1}}
\providecommand{\href}[2]{#2}
\providecommand{\path}[1]{#1}
\providecommand{\DOIprefix}{doi:}
\providecommand{\ArXivprefix}{arXiv:}
\providecommand{\URLprefix}{URL: }
\providecommand{\Pubmedprefix}{pmid:}
\providecommand{\doi}[1]{\href{http://dx.doi.org/#1}{\path{#1}}}
\providecommand{\Pubmed}[1]{\href{pmid:#1}{\path{#1}}}
\providecommand{\bibinfo}[2]{#2}
\ifx\xfnm\relax \def\xfnm[#1]{\unskip,\space#1}\fi
\bibitem[{Yuan and Gygi(2010)}]{Yuan2010}
\bibinfo{author}{G.~Yuan}, \bibinfo{author}{F.~Gygi},
  \bibinfo{journal}{Computational Science \& Discovery} \bibinfo{volume}{3}
  (\bibinfo{year}{2010}) \bibinfo{pages}{015004}.
\bibitem[{CECAM(2013)}]{CECAM}
\bibinfo{author}{CECAM}, \bibinfo{year}{2013}. \URLprefix
  \url{http://esvv.cecam.org}.
\bibitem[{Aulbur et~al.(1999)Aulbur, J\"onsson, and Wilkins}]{Aulbur1999}
\bibinfo{author}{W.~G. Aulbur}, \bibinfo{author}{L.~J\"onsson},
  \bibinfo{author}{J.~W. Wilkins}, \bibinfo{title}{Quasiparticle Calculations
  in Solids}, volume~\bibinfo{volume}{54} of \textit{\bibinfo{series}{Solid
  State Physics}}, \bibinfo{publisher}{Academic Press}, \bibinfo{year}{1999}.
\bibitem[{Georges et~al.(1996)Georges, Kotliar, Krauth, and
  Rozenberg}]{Georges1996}
\bibinfo{author}{A.~Georges}, \bibinfo{author}{G.~Kotliar},
  \bibinfo{author}{W.~Krauth}, \bibinfo{author}{M.~J. Rozenberg},
  \bibinfo{journal}{Rev. Mod. Phys.} \bibinfo{volume}{68}
  (\bibinfo{year}{1996}) \bibinfo{pages}{13--125}.
\bibitem[{Onida et~al.(2002)Onida, Reining, and Rubio}]{Onida2002}
\bibinfo{author}{G.~Onida}, \bibinfo{author}{L.~Reining},
  \bibinfo{author}{A.~Rubio}, \bibinfo{journal}{Rev. Mod. Phys.}
  \bibinfo{volume}{74} (\bibinfo{year}{2002}) \bibinfo{pages}{601--659}.
\bibitem[{Cardona and Thewalt(2005)}]{Cardona2005a}
\bibinfo{author}{M.~Cardona}, \bibinfo{author}{M.~L.~W. Thewalt},
  \bibinfo{journal}{Rev. Mod. Phys.} \bibinfo{volume}{77}
  (\bibinfo{year}{2005}) \bibinfo{pages}{1173--1224}.
\bibitem[{Cardona(2005)}]{Cardona2005}
\bibinfo{author}{M.~Cardona}, \bibinfo{journal}{Solid State Communications}
  \bibinfo{volume}{133} (\bibinfo{year}{2005}) \bibinfo{pages}{3 -- 18}.
\bibitem[{Logothetidis et~al.(1992)Logothetidis, Petalas, Polatoglou, and
  Fuchs}]{Logothetidis1992}
\bibinfo{author}{S.~Logothetidis}, \bibinfo{author}{J.~Petalas},
  \bibinfo{author}{H.~M. Polatoglou}, \bibinfo{author}{D.~Fuchs},
  \bibinfo{journal}{Phys. Rev. B} \bibinfo{volume}{46} (\bibinfo{year}{1992})
  \bibinfo{pages}{4483--4494}.
\bibitem[{Ram\'irez et~al.(2006)Ram\'irez, Herrero, and
  Hern\'andez}]{Ramirez2006}
\bibinfo{author}{R.~Ram\'irez}, \bibinfo{author}{C.~P. Herrero},
  \bibinfo{author}{E.~R. Hern\'andez}, \bibinfo{journal}{Phys. Rev. B}
  \bibinfo{volume}{73} (\bibinfo{year}{2006}) \bibinfo{pages}{245202}.
\bibitem[{Giustino et~al.(2010)Giustino, Louie, and Cohen}]{Giustino2010}
\bibinfo{author}{F.~Giustino}, \bibinfo{author}{S.~G. Louie},
  \bibinfo{author}{M.~L. Cohen}, \bibinfo{journal}{Phys. Rev. Lett.}
  \bibinfo{volume}{105} (\bibinfo{year}{2010}) \bibinfo{pages}{265501}.
\bibitem[{Allen and Heine(1976)}]{Allen1976}
\bibinfo{author}{P.~B. Allen}, \bibinfo{author}{V.~Heine},
  \bibinfo{journal}{Journal of Physics C: Solid State Physics}
  \bibinfo{volume}{9} (\bibinfo{year}{1976}) \bibinfo{pages}{2305}.
\bibitem[{Allen and Cardona(1981)}]{Allen1981}
\bibinfo{author}{P.~B. Allen}, \bibinfo{author}{M.~Cardona},
  \bibinfo{journal}{Phys. Rev. B} \bibinfo{volume}{23} (\bibinfo{year}{1981})
  \bibinfo{pages}{1495--1505}.
\bibitem[{Ceperley and Alder(1980)}]{Ceperley1980}
\bibinfo{author}{D.~M. Ceperley}, \bibinfo{author}{B.~J. Alder},
  \bibinfo{journal}{Phys. Rev. Lett.} \bibinfo{volume}{45}
  (\bibinfo{year}{1980}) \bibinfo{pages}{566--569}.
\bibitem[{Perdew and Zunger(1981)}]{Perdew1981}
\bibinfo{author}{J.~P. Perdew}, \bibinfo{author}{A.~Zunger},
  \bibinfo{journal}{Phys. Rev. B} \bibinfo{volume}{23} (\bibinfo{year}{1981})
  \bibinfo{pages}{5048--5079}.
\bibitem[{Martin(2004)}]{Martin2004}
\bibinfo{author}{R.~M. Martin}, \bibinfo{title}{Electronic Structure. Basic
  Theory and Practical Methods}, \bibinfo{publisher}{Cambridge University
  Press}, \bibinfo{year}{2004}.
\bibitem[{Gonze et~al.(2011)Gonze, Boulanger, and C\^ot\'e}]{Gonze2011}
\bibinfo{author}{X.~Gonze}, \bibinfo{author}{P.~Boulanger},
  \bibinfo{author}{M.~C\^ot\'e}, \bibinfo{journal}{Annalen der Physik}
  \bibinfo{volume}{523} (\bibinfo{year}{2011}) \bibinfo{pages}{168}.
\bibitem[{Gonze et~al.(2009)Gonze, Amadon, Anglade, Beuken, Bottin, Boulanger,
  Bruneval, Caliste, Caracas, Côté, Deutsch, Genovese, Ghosez, Giantomassi,
  Goedecker, Hamann, Hermet, Jollet, Jomard, Leroux, Mancini, Mazevet,
  Oliveira, Onida, Pouillon, Rangel, Rignanese, Sangalli, Shaltaf, Torrent,
  Verstraete, Zerah, and Zwanziger}]{Gonze2009}
\bibinfo{author}{X.~Gonze}, \bibinfo{author}{B.~Amadon}, \bibinfo{author}{P.-M.
  Anglade}, \bibinfo{author}{J.-M. Beuken}, \bibinfo{author}{F.~Bottin},
  \bibinfo{author}{P.~Boulanger}, \bibinfo{author}{F.~Bruneval},
  \bibinfo{author}{D.~Caliste}, \bibinfo{author}{R.~Caracas},
  \bibinfo{author}{M.~Côté}, \bibinfo{author}{T.~Deutsch},
  \bibinfo{author}{L.~Genovese}, \bibinfo{author}{P.~Ghosez},
  \bibinfo{author}{M.~Giantomassi}, \bibinfo{author}{S.~Goedecker},
  \bibinfo{author}{D.~Hamann}, \bibinfo{author}{P.~Hermet},
  \bibinfo{author}{F.~Jollet}, \bibinfo{author}{G.~Jomard},
  \bibinfo{author}{S.~Leroux}, \bibinfo{author}{M.~Mancini},
  \bibinfo{author}{S.~Mazevet}, \bibinfo{author}{M.~Oliveira},
  \bibinfo{author}{G.~Onida}, \bibinfo{author}{Y.~Pouillon},
  \bibinfo{author}{T.~Rangel}, \bibinfo{author}{G.-M. Rignanese},
  \bibinfo{author}{D.~Sangalli}, \bibinfo{author}{R.~Shaltaf},
  \bibinfo{author}{M.~Torrent}, \bibinfo{author}{M.~Verstraete},
  \bibinfo{author}{G.~Zerah}, \bibinfo{author}{J.~Zwanziger},
  \bibinfo{journal}{Computer Physics Communications} \bibinfo{volume}{180}
  (\bibinfo{year}{2009}) \bibinfo{pages}{2582 -- 2615}.
\bibitem[{Marini et~al.(2009)Marini, Hogan, Gr\"uning, and
  Varsano}]{Marini2009}
\bibinfo{author}{A.~Marini}, \bibinfo{author}{C.~Hogan},
  \bibinfo{author}{M.~Gr\"uning}, \bibinfo{author}{D.~Varsano},
  \bibinfo{journal}{Computer Physics Communications} \bibinfo{volume}{180}
  (\bibinfo{year}{2009}) \bibinfo{pages}{1392 -- 1403}.
\bibitem[{Giannozzi et~al.(2009)Giannozzi, Baroni, Bonini, Calandra, Car,
  Cavazzoni, Ceresoli, Chiarotti, Cococcioni, Dabo, Corso, de~Gironcoli,
  Fabris, Fratesi, Gebauer, Gerstmann, Gougoussis, Kokalj, Lazzeri,
  Martin-Samos, Marzari, Mauri, Mazzarello, Paolini, Pasquarello, Paulatto,
  Sbraccia, Scandolo, Sclauzero, Seitsonen, Smogunov, Umari, and
  Wentzcovitch}]{Giannozzi2009}
\bibinfo{author}{P.~Giannozzi}, \bibinfo{author}{S.~Baroni},
  \bibinfo{author}{N.~Bonini}, \bibinfo{author}{M.~Calandra},
  \bibinfo{author}{R.~Car}, \bibinfo{author}{C.~Cavazzoni},
  \bibinfo{author}{D.~Ceresoli}, \bibinfo{author}{G.~L. Chiarotti},
  \bibinfo{author}{M.~Cococcioni}, \bibinfo{author}{I.~Dabo},
  \bibinfo{author}{A.~D. Corso}, \bibinfo{author}{S.~de~Gironcoli},
  \bibinfo{author}{S.~Fabris}, \bibinfo{author}{G.~Fratesi},
  \bibinfo{author}{R.~Gebauer}, \bibinfo{author}{U.~Gerstmann},
  \bibinfo{author}{C.~Gougoussis}, \bibinfo{author}{A.~Kokalj},
  \bibinfo{author}{M.~Lazzeri}, \bibinfo{author}{L.~Martin-Samos},
  \bibinfo{author}{N.~Marzari}, \bibinfo{author}{F.~Mauri},
  \bibinfo{author}{R.~Mazzarello}, \bibinfo{author}{S.~Paolini},
  \bibinfo{author}{A.~Pasquarello}, \bibinfo{author}{L.~Paulatto},
  \bibinfo{author}{C.~Sbraccia}, \bibinfo{author}{S.~Scandolo},
  \bibinfo{author}{G.~Sclauzero}, \bibinfo{author}{A.~P. Seitsonen},
  \bibinfo{author}{A.~Smogunov}, \bibinfo{author}{P.~Umari},
  \bibinfo{author}{R.~M. Wentzcovitch}, \bibinfo{journal}{Journal of Physics:
  Condensed Matter} \bibinfo{volume}{21} (\bibinfo{year}{2009})
  \bibinfo{pages}{395502}.
\bibitem[{Marini(2008)}]{Marini2008}
\bibinfo{author}{A.~Marini}, \bibinfo{journal}{Phys. Rev. Lett.}
  \bibinfo{volume}{101} (\bibinfo{year}{2008}) \bibinfo{pages}{106405}.
\bibitem[{Cannuccia and Marini(2011)}]{Cannuccia2011}
\bibinfo{author}{E.~Cannuccia}, \bibinfo{author}{A.~Marini},
  \bibinfo{journal}{Phys. Rev. Lett.} \bibinfo{volume}{107}
  (\bibinfo{year}{2011}) \bibinfo{pages}{255501}.
\bibitem[{Cannuccia and Marini(2012)}]{Cannuccia2012}
\bibinfo{author}{E.~Cannuccia}, \bibinfo{author}{A.~Marini},
  \bibinfo{journal}{The European Physical Journal B} \bibinfo{volume}{85}
  (\bibinfo{year}{2012}) \bibinfo{pages}{1--7}.
\bibitem[{Cannuccia and Marini(2013)}]{Cannuccia2013}
\bibinfo{author}{E.~Cannuccia}, \bibinfo{author}{A.~Marini},
  \bibinfo{journal}{arXiv: cond-mat.mtrl-sci 1304:0072}
  (\bibinfo{year}{2013}).
\bibitem[{Gonze(1997)}]{Gonze1997}
\bibinfo{author}{X.~Gonze}, \bibinfo{journal}{Phys. Rev. B}
  \bibinfo{volume}{55} (\bibinfo{year}{1997}) \bibinfo{pages}{10337--10354}.
\bibitem[{Baroni et~al.(1987)Baroni, Giannozzi, and Testa}]{Baroni1987}
\bibinfo{author}{S.~Baroni}, \bibinfo{author}{P.~Giannozzi},
  \bibinfo{author}{A.~Testa}, \bibinfo{journal}{Phys. Rev. Lett.}
  \bibinfo{volume}{58} (\bibinfo{year}{1987}) \bibinfo{pages}{1861--1864}.
\bibitem[{Pavone et~al.(1993)Pavone, Karch, Sch\"utt, Strauch, Windl,
  Giannozzi, and Baroni}]{Pavone1993}
\bibinfo{author}{P.~Pavone}, \bibinfo{author}{K.~Karch},
  \bibinfo{author}{O.~Sch\"utt}, \bibinfo{author}{D.~Strauch},
  \bibinfo{author}{W.~Windl}, \bibinfo{author}{P.~Giannozzi},
  \bibinfo{author}{S.~Baroni}, \bibinfo{journal}{Phys. Rev. B}
  \bibinfo{volume}{48} (\bibinfo{year}{1993}) \bibinfo{pages}{3156--3163}.
\bibitem[{Baroni et~al.(2001)Baroni, de~Gironcoli, Dal~Corso, and
  Giannozzi}]{Baroni2001}
\bibinfo{author}{S.~Baroni}, \bibinfo{author}{S.~de~Gironcoli},
  \bibinfo{author}{A.~Dal~Corso}, \bibinfo{author}{P.~Giannozzi},
  \bibinfo{journal}{Rev. Mod. Phys.} \bibinfo{volume}{73}
  (\bibinfo{year}{2001}) \bibinfo{pages}{515--562}.
\bibitem[{Gonze(2005)}]{Gonze2005}
\bibinfo{author}{X.~Gonze}, \bibinfo{journal}{Zeitschrift fur Kristallographie}
  \bibinfo{volume}{220} (\bibinfo{year}{2005}) \bibinfo{pages}{558--562}.
\bibitem[{Fan(1950)}]{Fan1950}
\bibinfo{author}{H.~Y. Fan}, \bibinfo{journal}{Phys. Rev.} \bibinfo{volume}{78}
  (\bibinfo{year}{1950}) \bibinfo{pages}{808--809}.
\bibitem[{Fan(1951)}]{Fan1951}
\bibinfo{author}{H.~Y. Fan}, \bibinfo{journal}{Phys. Rev.} \bibinfo{volume}{82}
  (\bibinfo{year}{1951}) \bibinfo{pages}{900--905}.
\bibitem[{Sternheimer(1954)}]{Sternheimer1954}
\bibinfo{author}{R.~M. Sternheimer}, \bibinfo{journal}{Phys. Rev.}
  \bibinfo{volume}{96} (\bibinfo{year}{1954}) \bibinfo{pages}{951--968}.
\bibitem[{Hohenberg and Kohn(1964)}]{Hohenberg1964}
\bibinfo{author}{P.~Hohenberg}, \bibinfo{author}{W.~Kohn},
  \bibinfo{journal}{Phys. Rev.} \bibinfo{volume}{136} (\bibinfo{year}{1964})
  \bibinfo{pages}{B864--B871}.
\bibitem[{Kohn and Sham(1965)}]{Kohn1965}
\bibinfo{author}{W.~Kohn}, \bibinfo{author}{L.~J. Sham},
  \bibinfo{journal}{Phys. Rev.} \bibinfo{volume}{140} (\bibinfo{year}{1965})
  \bibinfo{pages}{A1133--A1138}.
\bibitem[{Troullier and Martins(1991)}]{Troullier1991}
\bibinfo{author}{N.~Troullier}, \bibinfo{author}{J.~L. Martins},
  \bibinfo{journal}{Phys. Rev. B} \bibinfo{volume}{43} (\bibinfo{year}{1991})
  \bibinfo{pages}{1993--2006}.
\bibitem[{Fuchs and Scheffler(1999)}]{Fuchs1999}
\bibinfo{author}{M.~Fuchs}, \bibinfo{author}{M.~Scheffler},
  \bibinfo{journal}{Computer Physics Communications} \bibinfo{volume}{119}
  (\bibinfo{year}{1999}) \bibinfo{pages}{67 -- 98}.
\bibitem[{Monkhorst and Pack(1976)}]{Monkhorst1976}
\bibinfo{author}{H.~J. Monkhorst}, \bibinfo{author}{J.~D. Pack},
  \bibinfo{journal}{Phys. Rev. B} \bibinfo{volume}{13} (\bibinfo{year}{1976})
  \bibinfo{pages}{5188--5192}.
\bibitem[{Perdew and Wang(1992)}]{Perdew1992}
\bibinfo{author}{J.~P. Perdew}, \bibinfo{author}{Y.~Wang},
  \bibinfo{journal}{Phys. Rev. B} \bibinfo{volume}{45} (\bibinfo{year}{1992})
  \bibinfo{pages}{13244--13249}.
\bibitem[{Perdew et~al.(1996)Perdew, Burke, and Ernzerhof}]{Perdew1996}
\bibinfo{author}{J.~P. Perdew}, \bibinfo{author}{K.~Burke},
  \bibinfo{author}{M.~Ernzerhof}, \bibinfo{journal}{Phys. Rev. Lett.}
  \bibinfo{volume}{77} (\bibinfo{year}{1996}) \bibinfo{pages}{3865--3868}.
\bibitem[{Hartwigsen et~al.(1998)Hartwigsen, Goedecker, and
  Hutter}]{Hartwigsen1998}
\bibinfo{author}{C.~Hartwigsen}, \bibinfo{author}{S.~Goedecker},
  \bibinfo{author}{J.~Hutter}, \bibinfo{journal}{Phys. Rev. B}
  \bibinfo{volume}{58} (\bibinfo{year}{1998}) \bibinfo{pages}{3641--3662}.
\bibitem[{Gonze and Lee(1997)}]{Gonze1997a}
\bibinfo{author}{X.~Gonze}, \bibinfo{author}{C.~Lee}, \bibinfo{journal}{Phys.
  Rev. B} \bibinfo{volume}{55} (\bibinfo{year}{1997})
  \bibinfo{pages}{10355--10368}.
\bibitem[{Zollner et~al.(1992)Zollner, Cardona, and Gopalan}]{Zollner1992}
\bibinfo{author}{S.~Zollner}, \bibinfo{author}{M.~Cardona},
  \bibinfo{author}{S.~Gopalan}, \bibinfo{journal}{Phys. Rev. B}
  \bibinfo{volume}{45} (\bibinfo{year}{1992}) \bibinfo{pages}{3376--3385}.

\end{thebibliography}

%
\end{document}